\documentclass[]{aastex63}

\newcommand{\fracbrac}[2]{\left(\frac{#1}{#2}\right)}
\newcommand{\dd}[2]{\frac{d#1}{d#2}}

\newcommand{\paren}[1]{\left(#1\right)}

\usepackage{natbib}
\usepackage{graphicx}
\usepackage{amsmath}
\usepackage{multirow}
\usepackage{ulem}

\accepted{Apr 17, 2022}
\submitjournal{APJ}

\shorttitle{Resonance Splitting}
\shortauthors{Murray et al.}
\graphicspath{{./}{}}

\begin{document}

\title{The Effects of Disk Induced Apsidal Precession on Planets Captured into Mean Motion Resonance}

\correspondingauthor{Zach~Murray}
\email{zachary.murray@cfa.harvard.edu}
\author[0000-0002-8076-3854]{Zachary~Murray}
\affiliation{Harvard-Smithsonian Center for Astrophysics, 60 Garden St., MS 51, Cambridge, MA 02138, USA}
\author[0000-0002-1032-0783]{Sam~Hadden}
\affiliation{Canadian Institute for Theoretical Astrophysics, 60 St George St Toronto, ON M5S 3H8, Canada}
\affiliation{Harvard-Smithsonian Center for Astrophysics, 60 Garden St., MS 51, Cambridge, MA 02138, USA}
\author[0000-0002-1139-4880]{Matthew~J.~Holman}
\affiliation{Harvard-Smithsonian Center for Astrophysics, 60 Garden St., MS 51, Cambridge, MA 02138, USA}

\begin{abstract}
    The process of migration into resonance capture has been well studied for planetary systems where the gravitational potential is generated exclusively by the star and planets. However, massive protoplanetary disks add a significant perturbation to these models. In this paper we consider two limiting cases of disk-induced precession on migrating planets and find that small amounts of precession significantly affect the equilibrium reached by migrating planets.  We investigate these effects with a combination of semi-analytic models of the resonance and numerical integrations.  We also consider the case of the disk's dispersal, which can excite significant libration amplitude and can cause ejection from resonance for large enough precession rates. Both of these effects have implications for interpreting the known exoplanet population and may prove to be important considerations as the population of well-characterized exoplanet systems continues to grow. 
    
\end{abstract}


\section{Introduction} 
\label{sec:intro}

     The capture of migrating bodies around a dominant central mass into mean motion resonances is a well-studied phenomenon.  Early studies explored resonance capture among satellites of the solar system giant planets subject to tidal migration \citep[e.g.,][]{Goldreich.1965,Yoder_1973,Henrard1983,Tittemore_1988}.
     Since these early studies, numerous resonant or near-resonant exoplanet systems have been discovered with both the radial velocity (RV) 
     \citep[e.g.,][]{Marcy2001,JohnSon_2011,Wright_2011}
     and transit method
     \citep[e.g.,][]{Mills_2016,MacDonald_2016,Luger_2017}.
     These systems have prompted studies of resonance capture in a planetary context ~\citep[e.g.,][]{Beauge_2006,Mustill_2011,Deck_2015} where gravitational interactions with the protoplanetary disk can drive migration and capture \citep{GT80,KleyNelson2012}. 
     This migration and capture is reproduced in hydrodynamic simulations \citep[e.g.,][]{Masset_2001,Laughlin_2002,Kley_2004,Rein2010} and models that include migration and eccentricity damping forces meant to mimic interactions within a protoplanetary disk can reproduce the orbital configurations of observed systems
     ~\citep[e.g.,][]{Lee_2002,Delisle_2017,Hadden2020}. 

     Resonant exoplanet system's present-day orbital configurations can serve as indirect probes of the natal disk conditions under which the planets were captured into resonance.   While recent initiatives like DSHARP have observed the large scale structures of protoplanetary disks \citep{Andrews_2018}, except for a few of the closest disks~\citep[e.g.][]{Andrews_2016}, the properties of the central few AU have not generally been observationally accessible. Currently, and for the foreseeable future, only indirect methods can probe the inner disk.  This motivates our investigation, as described below. 
    
     Traditional migration theory posits that the eccentricities of planets migrating into resonance are set by the ratio of their convergent migration rate to their eccentricity damping timescale. In particular, planets reach eccentricities $e\sim \sqrt{\tau_e/\tau_a}$ where {$\tau_e^{-1} = d\ln e /dt$ and $\tau_a^{-1}  = d\ln a /dt$}  $\dd{e}{t} = -e/\tau_e $ and $\dd{a}{t} = -a/\tau_a$ are the rates of eccentricity damping and semi-major axis migration \citep[e.g.,][]{Deck_2015}.  In general, both planets will become eccentric, and their individual eccentricities are such that the system resides in an equilibrium configuration. The equilibrium configuration depends on the ratio of the two planets' masses, with the less massive planet generally being more eccentric. However, traditional treatments of migration and capture usually neglect the influence of the disk on the equilibrium configuration reached by the planets. If the disk is sufficiently massive, its gravitational potential will induce periapsis precession that, as we show below, could alter the equilibrium eccentricities reached by a pair of migrating planets.  Previous work by \citet{Marzari_2018} explored how this disk potential shifts the semi-major ratio at which mean motion resonances occur.
     Whereas \citet{Marzari_2018} computes disk-induced resonance shifts by fitting mean period ratios of resonant planet pairs in ensembles of numerical simulations, we examine in detail how disk effects influence the dynamics of the resonance capture process, focusing on how disk-induced apsidal precession influences the growth of planets' eccentricities.

    This paper is organized as follows.   In Section \ref{sec:dynmodel}, we derive a Hamiltonian model for a resonance in the presence of additional precession and investigate the outcome of resonance capture under these conditions. Our analytic theory predicts strong excitation in equilibrium eccentricities for sufficiently large differential precession rates. We also detail an axisymmetric model of a massive disk and derive expressions for the precession rate it induces, and consider the timescales involved in its dispersal. In Section \ref{sec:results} we examine the outcomes of $N$-body simulations of resonant capture with an additional source of precession. We discuss the implications of these results in Section \ref{sec:discussion}.  We conclude in Section \ref{sec:summary} and describe future research directions.

\section{An Analytic Model for Resonance Capture with Apsidal Prescession}
\label{sec:dynmodel}

    In this section, we present Hamiltonian equations of motion that we use to model the dynamics of planets captured into resonance in the presence of a massive, precession-inducing disk. 
    We use this model to derive the equilibrium eccentricities reached by a pair of planets subject to migration and eccentricity damping forces.
    We show that, if a resonant planet pair's migration drives the inner planet into a disk cavity so that it expriences reduced eccentricity damping, its eccentricity can be significantly excited if the differential precession rate is large enough. 
    We derive an expression of the critical precession rate at which significant eccentricity can be excited. Finally, we discuss the validity and limitations of our simplified model.
    
    We consider the dynamics of a two planet system, with an inner planet of mass $m_1$ and outer planet of mass $m_2$ orbiting in or near a $j$:$j-1$ first-order mean motion resonance around a central star with mass $M_*$ and subject to an additional axis-symmetric external gravitational potential that induces apsidal precession at a rate $\dot\varpi_{i,\mathrm{add}}$ for the $i$th planet.\footnote{In general, an axis-symmetric potential will modify the mean motions of planets in addition to introducing apsidal precession. These modifications will influence the semi-major axis ratio at which a MMR occurs between planets. We ignore this effect in the simple Hamiltonian model presented in Equation \eqref{Hamiltonian:simplified} because it has little impact on planets' eccentricities, which is our main focus in this work.} 
    Following \citet{Hadden2019},  we adopt a Hamiltonian formalism and develop our equations of motion in terms of the canonical angle variables 
        $Q=j\lambda_2-(j-1)\lambda_1$,
            where $\lambda_i$ denotes the  mean longitude of the $i$th planet,
    and 
        $\gamma_i=-\varpi_i$ with $i=1,2$
            where $\varpi_i$ denotes the longitude of periapse of the $i$th planet,
    along with their conjugate action variables $P$ and $\Gamma'_i$. 
    The action variable $P$ is conjugate to the angle $Q$ and related to the planets period ratio, $P_2/P_1$, according to 
    \begin{equation}
        \frac{j-1}{j}\frac{P_2}{P_1} - 1
        =
        AP/j
    \end{equation}
    where $A\equiv\frac{3j}{2}\left(\frac{j}{\beta_2} + \frac{j-1}{\beta_1\sqrt{\alpha}}\right)$ with $\beta_i=m_i/(m_1+m_2)$ and $\alpha=a_1/a_2$. The action variables conjugate to $\gamma_i$ are
    $\Gamma_i' \approx \beta_i\sqrt{a_i}e_i^2$. We assume the planets are nearly coplanar and possess small eccentricities.  Thus, we truncate our equations of motion at first order in eccentricity and inclination. 
    Choosing units such that $\sqrt{G(M_* + m_2)/a_2^3} = 1$, the Hamiltonian of our system is given by 
    \begin{equation}
        H=-\frac{1}{2}A P^2 - 2\epsilon \paren{\tilde{f}\sqrt{\Gamma_1}\cos(Q + \gamma_1) + 
        \tilde{g}\sqrt{\Gamma_2}\cos(Q + \gamma_2)} -\dot\varpi_{1,\mathrm{add}}\Gamma_1 -\dot\varpi_{2,\mathrm{add}}\Gamma_2 
    \label{Hamiltonian:simplified}
    \end{equation}
    where $\epsilon \approx \frac{m_1 m_2}{M_* (m_1+m_2)}$, with $\tilde{f} = f \sqrt{\frac{m_1+m_2}{m_1 \sqrt{\alpha}}}$ and $\tilde{g}=g \sqrt{\frac{m_1+m_2}{m_2}}$ where $f$ and $g$ are order-unity constants, formulas for which are given in \citet{Hadden2019}. 

    To study the evolution of the system under the effects of migration and eccentricity damping induced by a disk, we augment Hamilton's equations by adding the following dissipative terms to the equations of motion
    \begin{eqnarray}
        \dd{\Gamma_i}{t}\bigg|_\mathrm{dis} &=& -2\frac{\Gamma_i}{\tau_{e,i}} \\
        \dd{P}{t}\bigg|_\mathrm{dis} &=& -\frac{3j}{2A}\left(1+ A P/j\right) \left(\frac{1}{\tau_{a,2}}-\frac{1}{\tau_{a,1}}\right)  \equiv 
        -\frac{3j}{2A} \frac{1}{\tau_{\alpha}} + \mathcal{O}(P) 
    \end{eqnarray}
     where $\dd{e_{i}}{t} = -e_{i}/\tau_{e,i}$ and $\dd{a_{i}}{t} =-a_{i}/\tau_{a,i} $ parameterize eccentricity damping and migration forces. 
    In the absence of dissipation, the quantity $D = \Gamma_1 + \Gamma_2 - P$ is conserved by Hamiltonian \eqref{eq:precham}.
    Under the effects of migration and eccentricity damping forces, the system will reach an equilibrium configuration that satisfies
    \begin{eqnarray}
    \dd{D}{t}\bigg|_\mathrm{dis} =-2\left(\frac{\Gamma_1}{\tau_{e,1}} + \frac{\Gamma_2}{\tau_{e,2}}\right) +\frac{3j}{2A\tau_\alpha}=0~.
   \label{eq:app:Ddot}
    \end{eqnarray}
    Provided the timescales $\tau_\alpha$ and $\tau_{e,i}$ are long compared to any other relevant dynamical timescales, the the equilibrium configuration reached by the system will be close to an equilibrium configuration of the conservative dynamics. At such an equilibrium, $Q + \gamma_1 = 0$ and $Q + \gamma_2 = \pi$, and Hamilton's equations imply
\begin{equation}
    \dd{}{t}(\gamma_1-\gamma_2)= -\epsilon\left(\frac{\tilde{f}}{\sqrt{\Gamma_1}} + \frac{\tilde{g}}{\sqrt{\Gamma_2}}\right) + \Delta\dot{\varpi}_\mathrm{add}=0~,
    \label{eq:delta_gamma_dot}
\end{equation}
    where $\Delta\dot{\varpi}_\mathrm{add} = \dot{\varpi}_\mathrm{2,add}-\dot{\varpi}_\mathrm{1,add}$. 
    Equations \eqref{eq:app:Ddot} and \eqref{eq:delta_gamma_dot} provide two equations for the two unknown values of $\Gamma_1$ and $\Gamma_2$ at equilibrium.
    Multiplying equation \eqref{eq:app:Ddot} by a factor of $\tau_\alpha$ and equation \eqref{eq:delta_gamma_dot} by a factor of $\epsilon^{-1}$, it is clear that the equilibrium eccentricities depend will depend on migration rates, eccentricity damping timescales, and apsidal precession rates through the parameter combinations $K_{i}\equiv{\tau_{\alpha}}/{\tau_{e,i}}$ and $\Delta\dot{\varpi}_\mathrm{add}/\epsilon$.

        While general solutions of Equations \eqref{eq:app:Ddot} and \eqref{eq:delta_gamma_dot} for $\Gamma_1$ and $\Gamma_2 $ involve roots of quartic polynomials, we can gain some intuition for the effect of the precession term, $\Delta\dot\varpi_\mathrm{add}$ by considering the limiting cases  
        $\Delta\dot\varpi_\mathrm{add}\ll\epsilon$  and
        $\Delta\dot\varpi_\mathrm{add}\gg \epsilon$.
        First, when $\Delta\dot\varpi_\mathrm{add}=0$, equilibrium occurs at 
        $\Gamma_\mathrm{1,eq0} = \frac{3j}{4K_\mathrm{eff}}{ \tilde f}^2$ 
        and 
        $\Gamma_\mathrm{2,eq0} = \frac{3j}{4K_\mathrm{eff}}{ \tilde g}^2$ 
        where  
        $K_\mathrm{eff} = A\left({ \tilde f}^2K_1+{\tilde g}^2K_2\right)$.  
        Rewriting these equilibrium values in terms of the planets' eccentricities, we obtain $e_\mathrm{1,eq0}=\frac{m_2+m_1}{m_1\sqrt{\alpha}}|f|\sqrt{\frac{3j}{4K_\mathrm{eff}}}$
        and 
        $e_\mathrm{2,eq0}=\frac{m_2+m_1}{m_2}g\sqrt{\frac{3j}{4K_\mathrm{eff}}}$.
        For a non-zero differential precession rate, the equilibrium $\Gamma_i$ values are shifted by an amount $\delta\Gamma_i$ with respect to the values $\Gamma_\mathrm{i,eq0}$.
        Equation \eqref{eq:app:Ddot} implies that these shifts are related to one another by $\delta \Gamma_2 =-\frac{\tau_{e,1}}{\tau_{e,2}}\delta\Gamma_1$.
        For $|\Delta\dot\varpi_\mathrm{add}|\ll\epsilon$, the shifts are given by
        $
        \delta \Gamma_1/\Gamma_\mathrm{1,eq0} 
        = 
        \sqrt{
        \frac
            {3j}
            {
            K_\mathrm{eff}
            }
        }
        \paren{
            1
            +
            \frac{K_1\Gamma_\mathrm{1,eq0}}{K_2\Gamma_\mathrm{2,eq0}}
            }^{-1}
        \frac
        {\Delta\dot\varpi_\mathrm{add}}
        {\epsilon}
        $
        and 
        $\delta \Gamma_2/\Gamma_\mathrm{2,eq0} 
        = 
        -\sqrt{
        \frac
            {3j}
            {
            K_\mathrm{eff}
            }
        }
        \paren{
            1
            +
            \frac{K_2\Gamma_\mathrm{2,eq0}}{K_2\Gamma_\mathrm{1,eq0}}
            }^{-1}
        \frac
        {\Delta\dot\varpi_\mathrm{add}}
        {\epsilon}
        $, to first order in  $\Delta\dot\varpi_\mathrm{add}/\epsilon$.
        Thus, a positive differential precession rate ($\Delta\dot\varpi_\mathrm{add}>0$) causes an increase in $e_1$ and a decrease in $e_2$ relative to the precession-free equilibrium values.
        When $\Delta\dot\varpi_\mathrm{add}\gg\epsilon$,
        $\Gamma_1 = \frac{3j}{4AK_1}-\tilde{g}^2 \frac{K_2}{K_1}\fracbrac{\epsilon}{\Delta\dot\varpi_\mathrm{add}}^2$ 
         and 
        $\Gamma_2 = \tilde{g}^2\fracbrac{\epsilon}{\Delta\dot\varpi_\mathrm{add}}^2$ 
        and when 
        $\Delta\dot\varpi_\mathrm{add}<0$ 
        and 
        $|\Delta\dot\varpi_\mathrm{add}|\gg\epsilon$,
        we find
        $\Gamma_1 = \tilde{f}^2\fracbrac{\epsilon}{\Delta\dot\varpi_\mathrm{add}}^2$
         and 
        $\Gamma_2 = \frac{3j}{4AK_2} - \tilde{f}^2\frac{K_1}{K_2}\fracbrac{\epsilon}{\Delta\dot\varpi_\mathrm{add}}^2$.
        These equilibrium values for $|\Delta\dot\varpi_\mathrm{add}|\gg\epsilon$ can be understood as the result of the precession terms appearing in Equation \eqref{eq:precham} splitting the first-order MMR into two distinct, well-separated resonances with resonant angles $Q+\gamma_1$ and $Q+\gamma_2$ occurring at $P\approx -\dot{\varpi}_{i,\mathrm{add}}/A$ for $i=1$ and 2, respectively. 
        When $\Delta\dot\varpi_\mathrm{add}>0$, the system captures reaches equilibrium in the $Q+\gamma_1$ resonance, the eccentricity of the inner planet is excited, and the equilibrium value is set principally by the ratio of $\tau_{1,e}/\tau_\alpha$. 
        Analogously, the dynamics are controlled by the outer planet's $Q+\gamma_2$ resonance when $\Delta\dot\varpi_\mathrm{add}<0$.

        If a pair of resonant planets migrating in a protoplanetary disk reach a location in the disk where density decreases rapidly, such as the disk inner edge or dead zone, the outer planet can push the inner one into the gap \citep[e.g.,][]{Ataiee2021}.
        In this scenario, the eccentricity damping effect of the disk on the inner planet should be greatly reduced.
        To determine the planet's equilibrium eccentricities in this scenario, we take 
        $\tau_{e,1}\rightarrow\infty$ 
        in Equation \eqref{eq:app:Ddot} and find 
        $e_{2,\mathrm{eq}} = \paren{2K_2\paren{j+(j-1)\frac{m_2}{m_1\sqrt{\alpha}}}}^{-1/2}$ 
        and 
    
    \begin{eqnarray}
         e_{1,\mathrm{eq}} = 
       \frac{m_2}{m_1\sqrt{\alpha}}
       \left|\frac{f}{g}\right|
       \left(\frac{1}{1- g^{-1}\fracbrac{\Delta\dot{\varpi}_{\mathrm{add}}}{n_2}\fracbrac{M_*}{m_1} e_{2,\mathrm{eq}}} \right)
       e_{2,\mathrm{eq}}~,
       \label{analytic:preds1}
    \end{eqnarray}
     where we now include the dependence on the outer planet's mean motion, $n_2$, 
     explicitly. 
     Equation \eqref{analytic:preds1} predicts that 
     $e_{1,\mathrm{eq}}$
     diverges when 
     $
     \Delta\dot{\varpi}_{\mathrm{add}} 
     = 
     n_2g\frac{m_1}{M_*}e_{2,\mathrm{eq}} = 
     n_2g\frac{m_1}{M_*}\paren{2K_2\paren{j+(j-1)\frac{m_2}{m_1\sqrt{\alpha}}}}^{1/2}
     $.
     While this divergence is an artifact of truncating our equations of motion at first order
     in eccentricities, numerical simulations presented below in Section \ref{sec:results} show that a large increase in the inner planet's equilibrium eccentricity does in fact occur when the differential precession approaches this critical rate.
     
\begin{figure}
        \centering
            \includegraphics[totalheight=7.5cm]{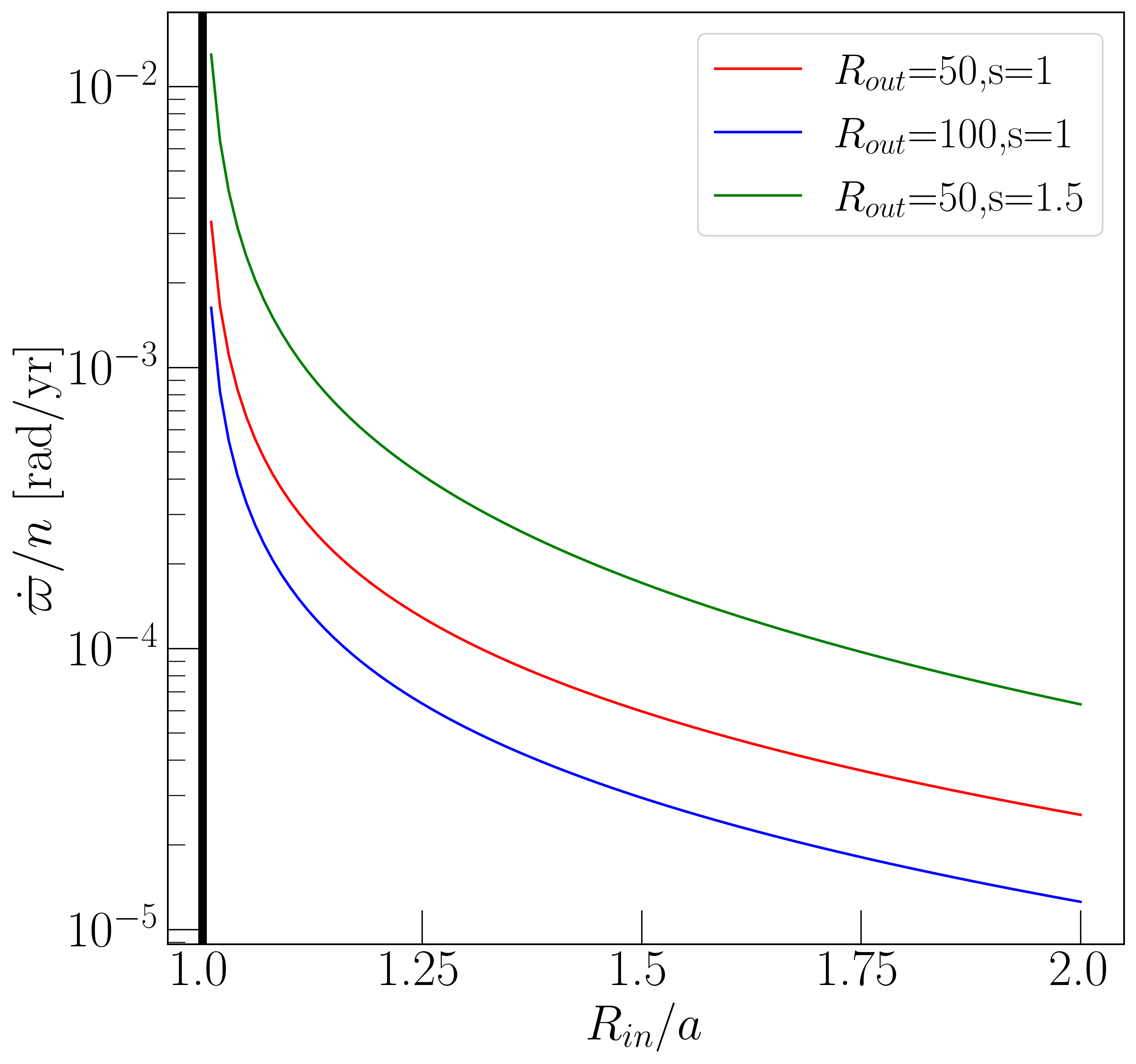}
            \caption{
            Precession rates induced by massive disk computed using Equation \eqref{eq:diskpotential} for different values of $s$ and $R_\mathrm{out}$. The disk mass was taken to be $M_\mathrm{disk} = 0.01 M_*$ and the planet was taken to have small eccentricity $e=0.01$ and semi major axis $a=1$ AU.
            } 
            \label{fig:diskprecession}
        \end{figure}

    Equipped with our analytic model, we now provide quantitative estimates of precession rates experienced by exoplanets in the central cavity of a massive, axisymmetric disk. To compute the apsidal precession induced by the protoplanetary disk, we closely follow \citet{Petrovich2019} and model the potential of the disk with a power-law surface density profile:
        \begin{equation}
        \Sigma(r) = \Sigma_0 \fracbrac{r}{R_\mathrm{in}}^{-s},
        \end{equation}
        where        
        \begin{equation}
        \Sigma_0 = \fracbrac{M_\mathrm{disk}}{2 \pi R_\mathrm{in}^2} \left[\frac{2-s}{\fracbrac{R_\mathrm{out}}{R_\mathrm{in}}^{2-s} -1}\right].
        \end{equation}
        Here, $R_\mathrm{in}$ is the radius of the inner edge of the central cavity, $R_\mathrm{out}$ the outer edge, $0 < s < 2$ is the power-law slope and $M_\mathrm{disk}$ the total mass of the disk. The potential generated by such a disk $\phi_\mathrm{disk}$ at a radial distance $r<R_\mathrm{in}$ is given by
        \begin{equation}
            \phi_\mathrm{disk}(r) = -\frac{1}{2}{G}
            \int_{R_\mathrm{in}}^{R_\mathrm{out}}\Sigma(R)b_{1/2}^{(0)}\left(\frac{r}{R}\right)dR~,
            \label{eq:diskpotential}
        \end{equation}
        Provided the timescales $\tau_\alpha$ and $\tau_{e,i}$ are long compared to any other relevant dynamical timescales, the the equilibrium configuration reached by the system will be close to an equilibrium configuration of the conservative dynamics.

    where  
        \begin{equation*}
            b_{s}^{(n)}(\alpha)= \frac{1}{\pi}\int_{-\pi}^\pi\frac{\cos(n\theta) d\theta}{(1+\alpha^2-2\alpha\cos\theta)^{s}}~.
        \end{equation*}
        is a Laplace coefficient \citep[e.g.,][]{MDbook}. We derive an expression for the orbit-averaged precession rate for a planet subject to the potential given by Equation \eqref{eq:diskpotential} as follows:
        first we substitute $r=a(1-e\cos u)$ in Equation \eqref{eq:diskpotential}, where $e, $ and $u$ ,  eccentricity, and eccentric anomaly of the planet. Next, we expand to second order in the planet's eccentricity. Finally we take the orbit average of the potential, $<\phi_\mathrm{disk}> = \int_{-\pi}^{\pi}\phi_\mathrm{disk}(r)\times(1-e\cos u)du$. Using Lagrange's planetary equations \citep{MDbook}, we derive the disk-induced precession rate
        \begin{equation}
            \frac{\dot\varpi}{n} = \eta a \frac{M_\mathrm{disk}}{M_*} 
            \int_{R_\mathrm{in}}^{R_\mathrm{out}} \fracbrac{a}{R}\fracbrac{R}{R_\mathrm{in}}^{-s} b_{3/2}^{(1)}\left(\frac{a}{R}\right) dR 
            \label{eq:precham}
        \end{equation}
        where $n=\sqrt{GM_*/a^{3}}$ is the planet's mean motion and $\eta = \frac{2-s}{4(R_\mathrm{in}^2\left((R_\mathrm{out}/R_\mathrm{in})^{2-s} -1\right)}$ is a normalization constant that depends on the disk size and power law slope. Figure \ref{fig:diskprecession} shows precession rate versus $R_\mathrm{in}/a$ for a few different disk surface densities. For $R_\mathrm{in}>>a$, $b_{3/2}^{(1)}\left(\frac{a}{R}\right)\approx 3 \frac{a}{R}$ and Equation \eqref{eq:precham} gives 
        \begin{equation}
            \dot\varpi \approx \frac{3 n}{4}\fracbrac{M_\mathrm{disk}}{M_*} \fracbrac{2-s}{1+s} \fracbrac{1-(R_\mathrm{in}/R_\mathrm{out})^{1+s}}{(R_\mathrm{out}/R_\mathrm{in})^{2-s}-1}\fracbrac{a}{R_\mathrm{in}}^3. 
        \end{equation}

     However, when $R_\mathrm{in}\sim a$, the induced precession rate increases steeply as the planet's orbit approaches the disk's inner edge. 
     
     The precession rate predicted by Equation \eqref{eq:precham} diverges when $a(1+e) = R_\mathrm{in}$, i.e., when the apoastron location of the outer planet is inside the disk.  When the planet is inside the disk, the planet's gravitational influence on the local mass distribution in the disk cannot be neglected.  \citet{Marzari_2016} compute apsidal precession rates of planets embedded in disks using hydrodynamical simulations and compare them to several analytic approximations~\citep[e.g.,][]{Binney_Tremaine,Mestel_1963,WARD_1981,Silsbee_2015}.  They found that if the planets embedded in the disk are very massive, their influence on the disk structure can result in a negative precession rate. Despite this complication, the model outlined in Section \ref{sec:dynmodel} can treat either case, as it makes no assumption about the sign of the precession rate. 
    
    Apsidal precession induced by a disk will modify the equilibrium eccentricities reached by a pair of planets that capture into resonance, as demonstrated in Section \ref{sec:dynmodel}.
     After the disk disperses, the equilibrium dynamical configuration will correspond to the conventional precessionless equilibrium.
    If the disk's dispersal is rapid compared to the secular interaction timescale of the planets, then planets will no longer be in equilibrium and instead exhibit oscillations in their eccentricities and resonant angles.  If dispersal is driven by photoevaporation - a process that removes material from the disk starting from an inner cavity of the disk and proceeds outwards, it can disperse on a timescale as short as $\approx 10^{5}$ years~\citep{Alexander_2006}. The exact time scale of photoevaporative dispersal is an open problem, but the steep dependence of precession rate on $a/{R_\mathrm{in}}$ (see Figure \ref{fig:diskprecession}) implies that the precession induced on the planets will decrease rapidly as the disk photoevaporates, since the majority of the precession induced on a planet comes from the portion of the disk closest to the planet. For example, for a disk where $R_\mathrm{out} = 100 \mathrm{AU}$ , that evaporates from the inside out in $10^5$ years, $R_\mathrm{out}$ increases by $\approx 10^{-3} \mathrm{AU/Yr}$.  If the system has an outer planet at $1 \mathrm{AU}$ and with the disk's inner edge at a few hill-radii from the outer planet $R_\mathrm{in} = 1.1 \mathrm{AU}$ initially, the precession rate will decline by $70 \%$ within $100$ years.  This rapid decrease will be even more significant if the planet starts closer to the disk.  In these situations the disk density should be decreasing over time, however, the resonant capture equilibrium will be set by the conditions in the disk shortly before its evaporation.  Hence, we do not need to model the history of the mass of the disk to study the dynamical consequences of it’s rapid dispersal.  In the case of a massive planets that generate a cavities in their disks, the precession rates will be more complicated than those suggested by \eqref{eq:precham}, however so long as the mode of dispersal is photoevaporation, the reduction in the precession rate will still be rapid. Therefore it is appropriate to consider the limit in which the bulk of disk's gravitational influence dissipates rapidly, in this paper we approximate the disk's dispersal as instantaneous.

\section{Results}
\label{sec:results}

    In this section, we compare our analytic predictions to $N$-body simulations and explore the different properties of the solutions.  All numerical integrations are done with the WHFast integrator \citep{whfast_2015} based on the symplectic mapping algorithm of \citet{Wisdom_1991} and implemented in the REBOUND code \citep{REBOUND}. In section \ref{sec:dynmodel} we predict that migrating planets can reach high eccentricites if there is a large difference in the precession rate between the planets.
    We therefore choose an integration time step set to $1/500$ of the inner planet's orbital period, ensuring the perihelion passage timescale, $P\sqrt{(1-e)^3/(1+e)}$, is resolved with 16 or more steps for planet eccentricities of $e_i < 0.875$ \citep{Wisdom_2015}.
    Additional eccentricity damping, migration, and periapsis precession effects are included in our simulations using the \texttt{modify\_orbits\_direct} routine of the REBOUNDx package \citep{REBOUNDx}.
    We set $\tau_{a,1}/\tau_{a,2} = -\frac{m_1}{m_2}\fracbrac{j}{j-1}^{2/3}$ in order to limit any bulk migration of the planet pairs in our simulations and study resonance capture outcomes at fixed values of $\dot\varpi_{2,\mathrm{add}}/{n_2}$.
    
    While our simulations maintain a constant $\dot\varpi_{i,\mathrm{add}}/{n_2}$ to focus on how differential precession influences the dynamics of resonant capture and the resulting post-capture equilibrium resonant state, the ratio $\dot\varpi_{i,\mathrm{add}}/n_2$ might continue to evolve if a planet pair continues to migrate after capture. Nevertheless, a resonant pair's dynamical state will simply track the evolving equilibrium configuration if the migration is not too rapid.

    For all of the simulations in this section, we examine motion near the 3:2 mean motion resonance, with equal mass planets taking $m_i = 5 \cdot 10^{-4}$ around a star with $M_* = 1 M_\odot$. We pick masses in the giant planet regime, similar to many of the observed resonant- and near-resonant planet pairs \citep{Wright_2011}.  Equation \eqref{analytic:preds1} predicts that the critical differential precession rate scales linearly with the inner planet's mass and we have confirmed this prediction holds with additional numerical simulations.   We generate a differential precession rate by imposing a nonzero $\dot\varpi_{2,\mathrm{add}}$ while keeping $\dot\varpi_{1,\mathrm{add}}$ zero.
    While in reality, both planets will be subject to apsidal precession caused by a disk's gravitational potential, the modified equilibrium eccentricities reached by the planet pair depends only on their differential precession, $\dot\varpi_{2,\mathrm{add}} -\dot\varpi_{1,\mathrm{add}}$.

\begin{figure}[!ht]
    \centering
        \includegraphics[totalheight=7.5cm]{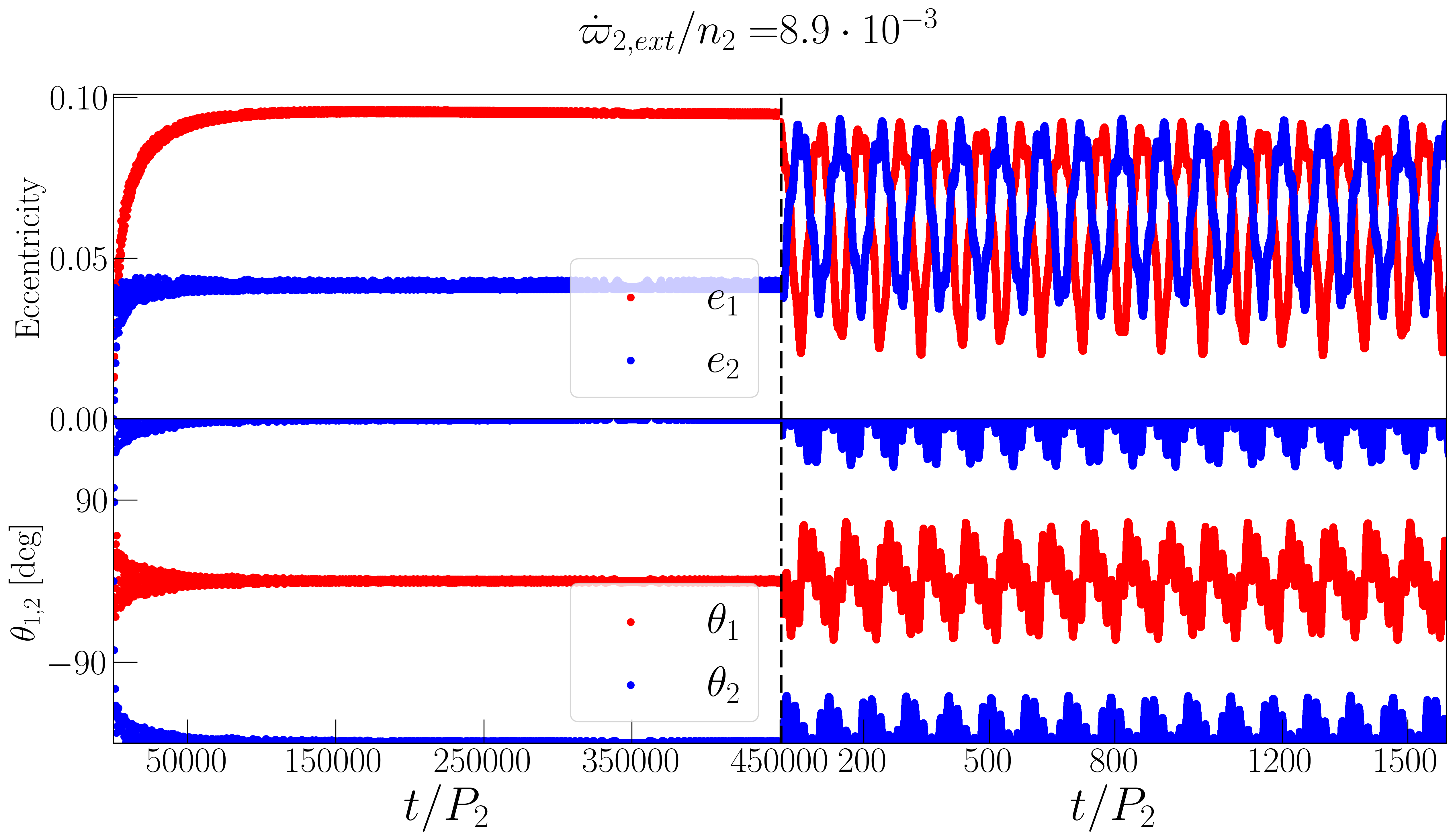}
        \caption{
        Results of a numerical simulation with migration timescale $\tau_{a,2} = 1.6 \times10^5 P_2$ , $K = 60$, and $\dot{\varpi}_{2,\mathrm{add}}=9\times10^{-3}n_2$. The upper left panel shows the evolution of the inner (red) and outer (blue) planets' eccentricities while the lower left portion shows the time evolution  resonant angles, $\theta_i = 3\lambda_2 - 2\lambda_1 - \varpi_i$.  
        The right panel shows the evolution of the system after the additional migration, eccentricity damping, and precession forces are suddenly removed. 
        The resulting mismatch between the perturbed and unperturbed equilibria result in an induced libration amplitude. }        
        \label{fig:convergedcase}
    \end{figure}

\begin{figure}[!ht]
    \centering
        \includegraphics[totalheight=7.5cm]{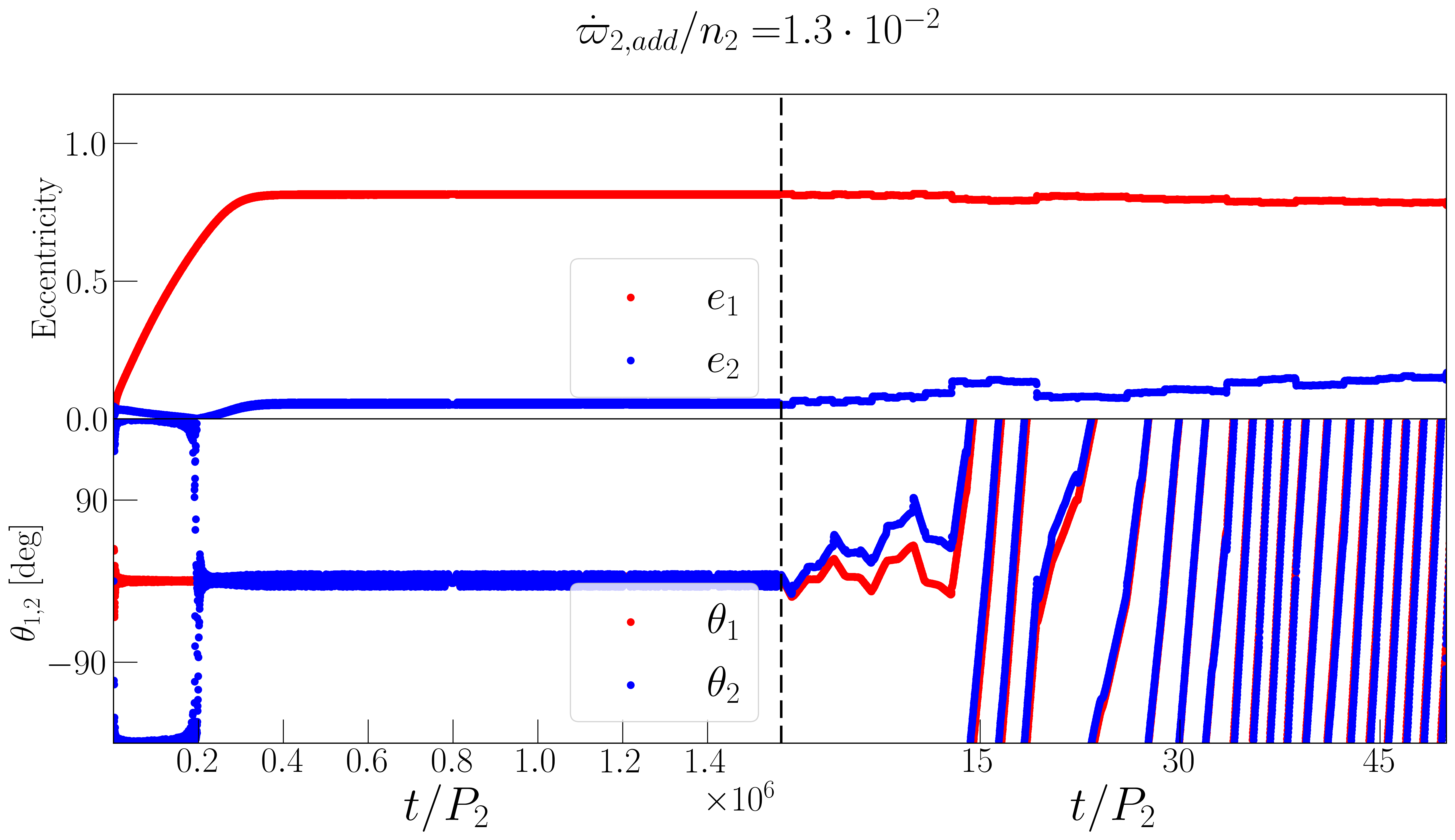}
        \caption{
        The effect of disk dispersal is illustrated here for a fiducial precession rate greater than the critical rate.  The left panel shows the evolution of the planets' eccentricities (red and blue) and corresponding resonant angles of the two planets over time.  The simulation runs for $0.3 \tau_{a,2}$ with $\tau_{a,2} = 10^6$ yrs and $K = 280$.  The right panel shows the resulting libration from disk dispersal on the much shorter timescale of a few thousand years. The vertical dashed line corresponds to the time at which the disk disperses, which we've taken to be an instantaneous process.  This example shows a change in the behavior of the resonant angles during capture ($\theta_2 - \theta_1 \neq \pi$).  Due to the planets reaching equilibrium at much higher eccentricity, the mismatch between the captured equilibrium is so large that stability is lost upon disk dispersal.}
        \label{fig:divergedcase}
    \end{figure}

    Figures \ref{fig:convergedcase} and \ref{fig:divergedcase} show the results of simulations of resonant capture with $K=280$ for two different $\dot\varpi_{2,\mathrm{add}}$ values. Figure \ref{fig:convergedcase} shows the capture and evolution of two planets with a differential precession rate less than the critical value.  The planets capture in resonance slightly away from the precession-free equilibrium, at relatively low eccentricities. 
    Figure \ref{fig:divergedcase} shows the capture and evolution of the same system but with a differential precession rate greater than the critical value.  The planets capture far away from the precession-free equilibrium and the inner body reaches high eccentricity.  There is also a qualitative change in the behavior resonant angles, and the equilibrium condition $\theta_2-\theta_1 \approx \pi$ is violated for captures with large differential precession. 

    After capture is complete, we turn off migration, eccentricity damping, and precession forces to mimic the rapid dispersal of the protoplanetary disk.    When precession effects are turned off, the migrating planets' eccentricities are no longer in equilibrium and begin to oscillate about new, precession-free equilibria. The resonant angles also begin to show oscillations after precession effects are turned off.  When the capture occurs with low differential precession, these oscillations are small, and the system retains its stability. When the forces are removed from a system with large differential precession, the resulting oscillations result in close encounters between the planets and loss of stability.

\begin{figure}[!ht]
    \centering
        \includegraphics[totalheight=7.5cm]{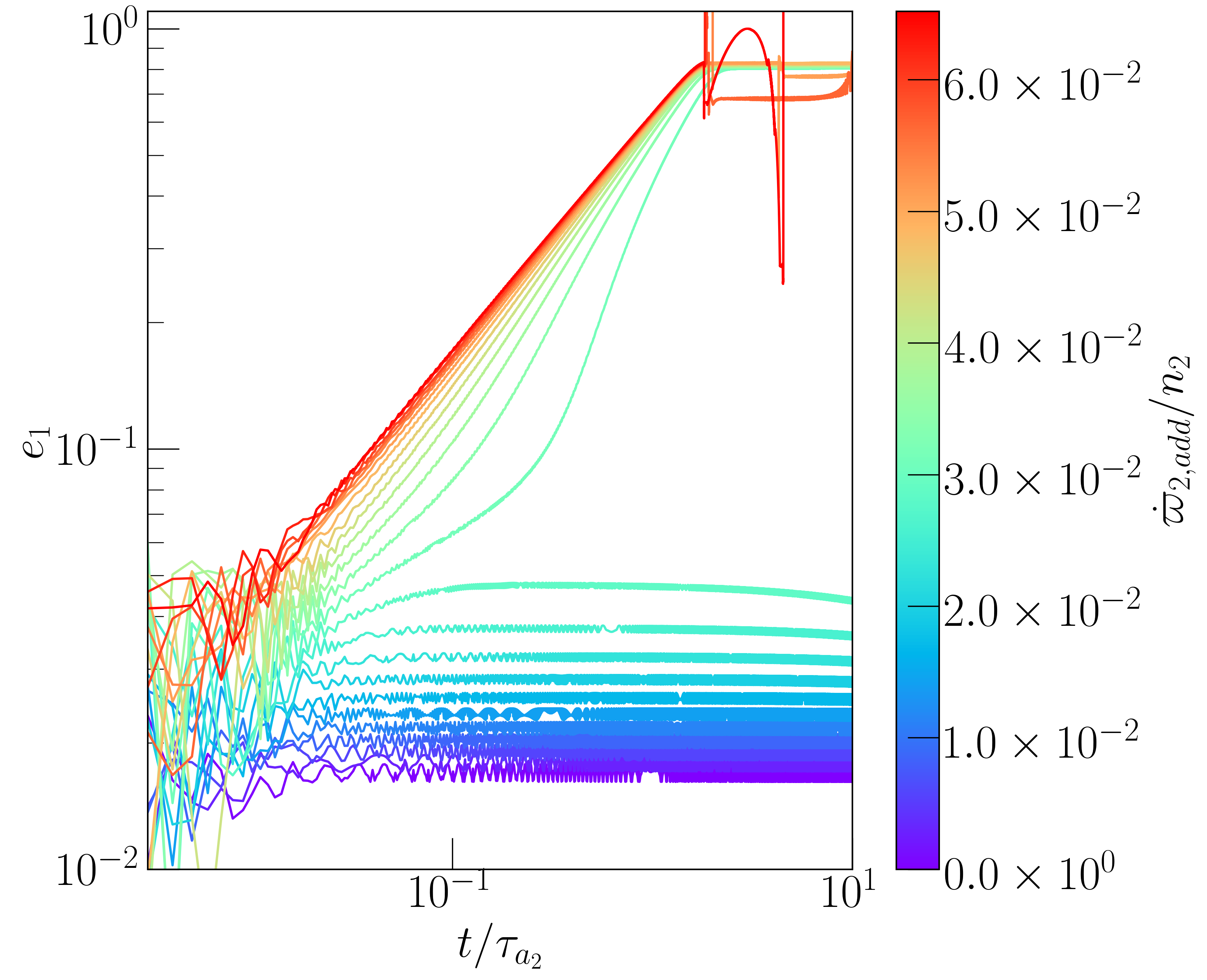}
        \caption{
        Time evolution of the inner planet's eccentricity, $e_1$ for a range of $ \dot\varpi_{2,\mathrm{add}}$ values.
        All simulations were run for $10\tau_{a_2}$ where $\tau_{a2} = 10^{6} \textit{yr} = 1.6 \cdot 10^5 P_2$. Planet masses were both set to $m_i = 3 \cdot 10^{-5}$ and eccentricity damping chosen so that $K=280$, which yields a critical precession rate of  $\dot\varpi_{2,\mathrm{add}} - \dot\varpi_{1,\mathrm{add}}  \approx 3 \cdot 10^{-3}$. 
        We can see that the eccentricity of system where resonance capture takes place with differential precession behaves in one of two main ways.  The first exists at low induced precession rates which results in the eccentricity of the inner body asymptotically low eccentricities. As the precession rate increases, the solutions quickly transition to saturating at high eccentricities and do so above a critical precession rate}
        \label{fig:e1vstime}
    \end{figure}

    Figure \ref{fig:e1vstime} shows simulation results for a range of $\dot\varpi_{2,\mathrm{add}}$ values assuming all of the damping in eccentricity was on the outer planet. The results illustrate $e_1$ can become large once a critical differential precession rate of the order $\sim (m_p/M_*)\times \sqrt{K}$ is reached, as predicted by the analytic model presented in Section \ref{sec:dynmodel}.  In contrast to the analytic model's prediction, our numerical simulations do not show that the equilibrium eccentricity decreases once this differential precession rate is passed. Instead, we find a sharp transition between solutions with low and high inner planet equilibrium eccentricities as a function of precession rate, and that systems with higher precession rates capture at increasingly higher eccentricities.  This trend continues until a critical precession rate is achieved, above which all systems begin to capture at much higher eccentricities. Note that not all planets in this second regime capture stably at high eccentricities, sufficiently large precession rates can result in instability in the resulting resonance, as can be seen in Figure \ref{fig:e1vstime}.  Additionally, the timescale associated with low and high eccentricity capture differ significantly, capture at low eccentricity occur within timescales $\approx 10^{-1} \tau_{P}$, whereas captures at high eccentricity take significantly longer. It may be possible - especially in the case of very gradual migration (and correspondingly large $\tau_{P}$) - that disk dispersal may occur in some systems before these equilibria are reached.  
    
\begin{figure}[!ht]
    \centering
        \includegraphics[totalheight=7.5cm]{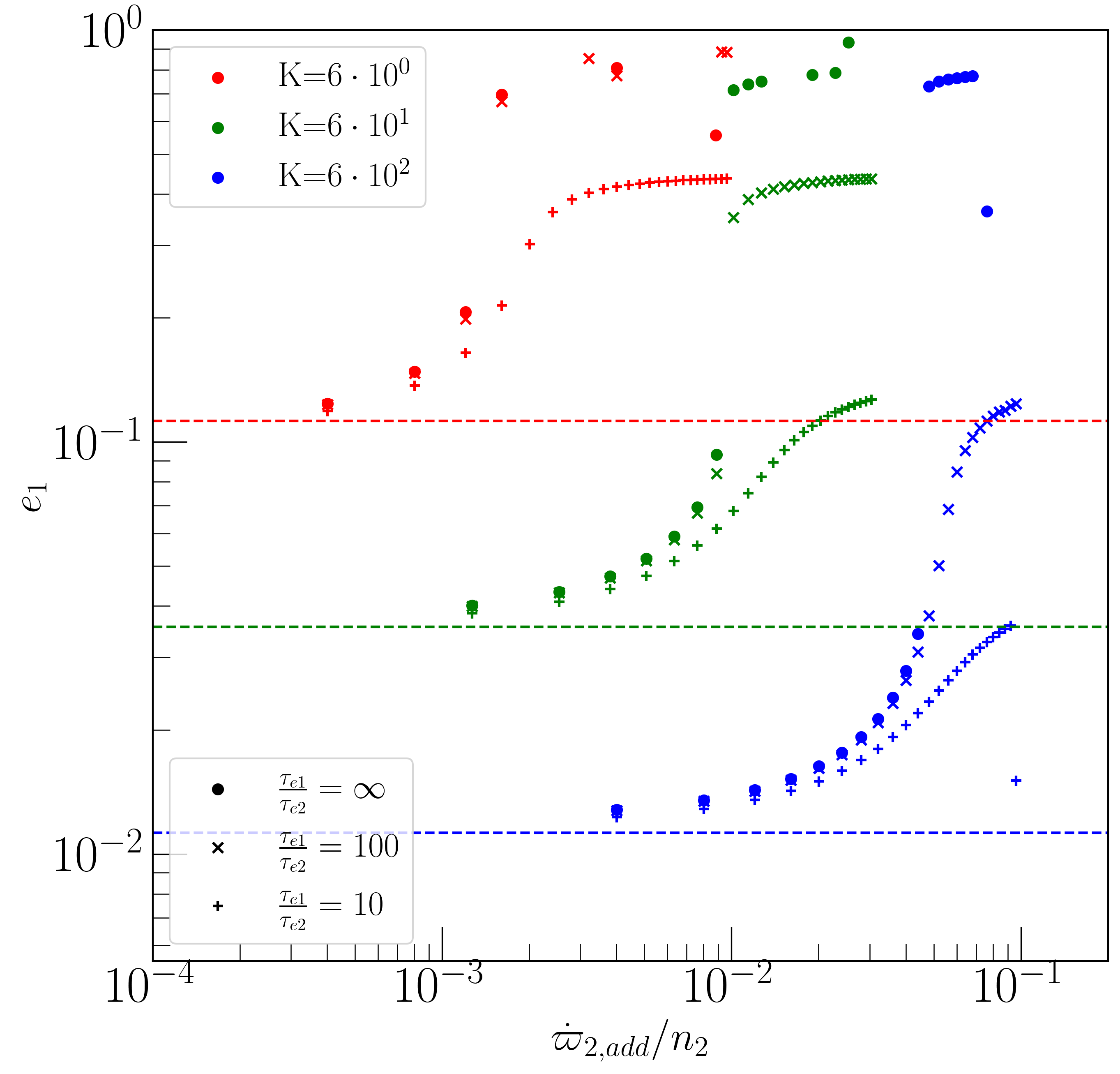} 
        \caption{The equilibrium eccentricities of the inner body as a function of $\dot{\varpi}_{2,\mathrm{add}}$ for several different $K$ and $\frac{\tau_{e1}}{\tau_{e2}}$ on the two planets. Equilibrium eccentricities are computed via numerical simulations that include extra migration, eccentricity damping, and prescession forces as described in the main text.  To determine the equilibrium eccentricities, numerical simulations were run for $10$ migration timescales, which was taken to be $\tau_{a,2} = 1.6 \cdot 10^5$ orbits of the outer planet.  The dashed lines are the equilibria predicted by Equation \eqref{analytic:preds1} with $\dot{\varpi}_{i,\mathrm{add}}=0$.
        } 
        \label{fig:predsvsk}
    \end{figure}

    Up to this point we have ignored any eccentricity damping experienced by the inner planet.   We relax this assumption in Figure \ref{fig:predsvsk}, where we show simulation outcomes over a range of precession rates for different inner planet eccentricity damping strengths.  Such a situation might occur if the inner planet is also embedded in the disk or could be due to tidal circularization from the host star.  Both of these effects will also cause migration of the inner planet, but so long as the migration rate is smaller than that of the outer planet resonant capture will still occur and the outcome of resonant capture will depend migration rates only via the combination $\tau_{\alpha} = (1/\tau_{a,2} - 1/\tau_{a,1})^{-1}$.  Figure \ref{fig:predsvsk} shows the equilibrium eccentricity, $e_{1,\mathrm{eq}}$, reached by the inner body as a function of $\varpi_{2,\mathrm{add}}$ for various values of $K$ and $\frac{\tau_{e1}}{\tau_{e2}}$, the ratio of the eccentricity damping timescales between the two planets.  We find a sharp transition between the low eccentricity and high eccentricity equilibria continues to exist over a wide variety of $K$ even when the damping on the inner planet is nonzero. It captures the $\approx \sqrt{K}$ scaling of the critical precession rate at $K > 50$ and provides an order of magnitude approximation at smaller $K$. We find that as the eccentricity damping on the inner body becomes larger, the transition between the two regimes becomes smoother. 

\begin{figure}
    \centering
        \includegraphics[totalheight=7.5cm]{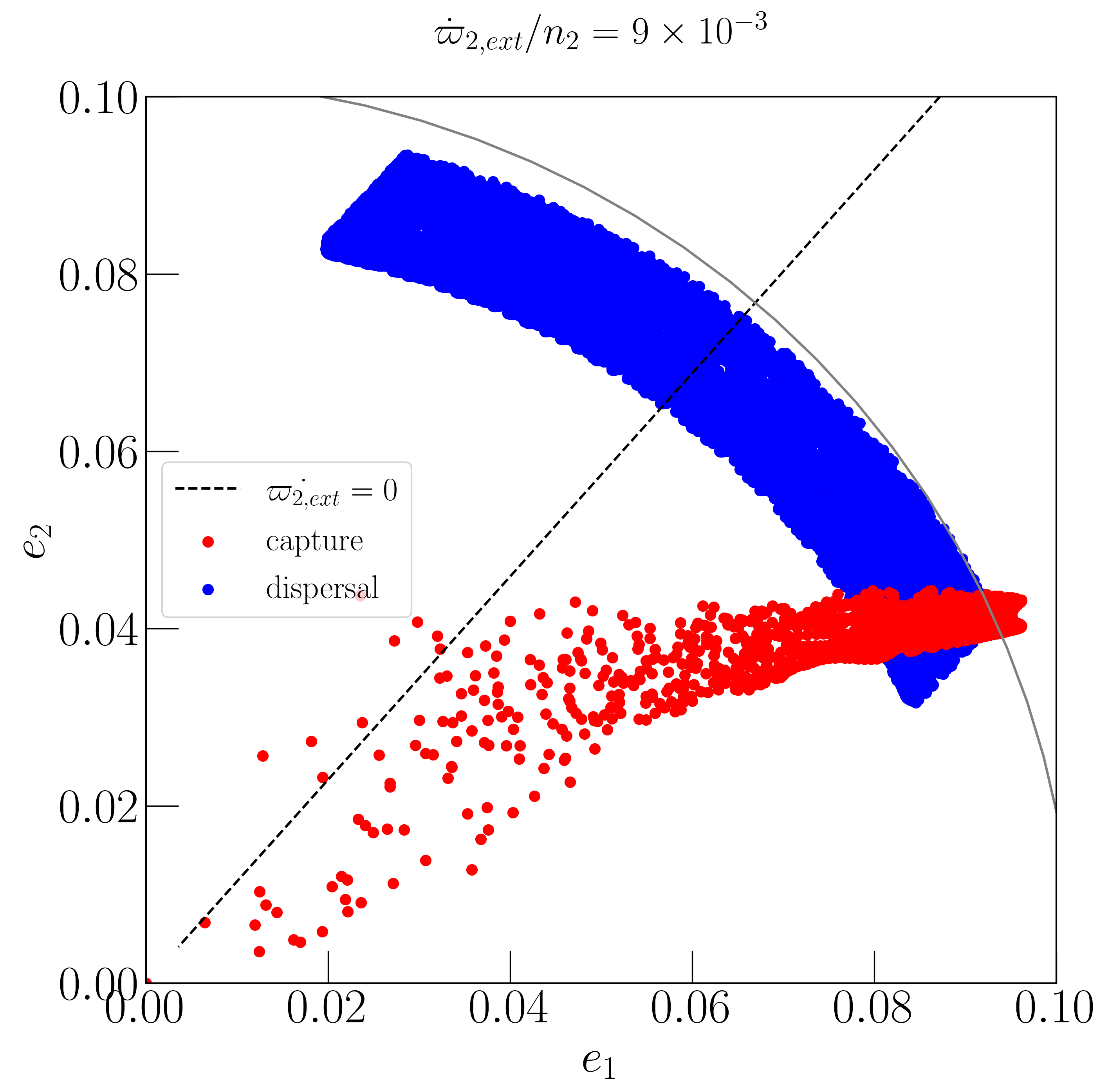}
            \caption{
            In the $e_1$ - $e_2$ plane resonant capture with additional precession results in deviation from the precession-less prediction (shown here as a dashed line).  The two planets start with nearly circular orbits and capture before migrating into resonance (red points) with $K=60$. After the disk disperses, the planets are left librating about the unperturbed equilibrium (blue points) with amplitude proportional to the distance from equilibrium.  The libration occurs about a line of approximately constant angular momenta (grey semi-circle).  This line is set by what would be expected of two non-interacting keplerian orbits, interactions between the planets cause them to deviate from this line. }
        \label{fig:deckwprec}
    \end{figure}

    The deviations in eccentricity from the precession-free case could be used to explain the structures of exoplanet systems.  Resonant capture without precession, as described in \cite{Deck_2015}, predicts a characteristic eccentricity ratio related to the mass of the planets and the captured resonance. As shown in Figure \ref{fig:deckwprec} including an additional source of precession allows planets to capture far from the predicted eccentricity ratio, which will induce a significant libration amplitude after disk dispersal.  This mechanism could be used to explain the origins of planetary systems that are found in resonance, but with significant libration amplitudes.  Conversely, as demonstrated by Figures \ref{fig:convergedcase} and \ref{fig:divergedcase}, the dynamics of resonant capture with an external source of precession predict a critical rate above which stability is lost upon disk dispersal.  Since surviving planets must have survived disk dispersal, the presence of the upper branch amounts to a constraint on the conditions in the planetary system at the time of capture and could be used to rule out sufficiently massive disks (or any other condition that imposes apsidal precession on the planets).
    
\section{Discussion}
\label{sec:discussion}

    In Section \ref{sec:results}  we showed that including additional precession results in capture at different equilibrium eccentricities from those where precession is neglected. The subsequent dispersal of the disk induces significant libration amplitudes in the captured planets. These libration amplitudes may be detectable in well-characterized systems.  Should measurements be sufficiently accurate to rule out large libration amplitudes, their absence can constrain disk induced precession rates during resonant capture. 

    When considering observability it is important to take into account the possibility that these amplitudes might be damped over Gyr timescales. Two potential pathways for damping libration amplitudes are tidal interactions and the ejection of smaller bodies.  First, stellar tides are expected to be effective in damping eccentricities of planets in older systems, but their strength falls off quickly with distance.  In many exoplanet systems \citep[e.g. for near-resonant Kepler planets, ][]{Lee_2013} tidal dissipation is too weak to change the eccentricities over Gyr timescales. 
    
    Second, and more uncertainly, damping might occur by ejecting smaller objects.  This mechanism must assume such a population of such objects of sufficient mass and proximity to the planets to damp the libration amplitudes.  Given the above considerations, we argue that it's possible that induced libration amplitudes will survive undamped for Gyr timescales and will therefore be observable. 
    
    We can use our expression in Section \ref{sec:dynmodel} to derive an approximate criterion to estimate how close a given system will be to the critical differential rate. We consider the case of an disk with power-law slope $s=1$ and $R_\mathrm{out} >> R_\mathrm{in}$.  We approximate the precession due to the disk as a power law, which is only a good approximation when $R_\mathrm{in}/a > 2$ and derive the following approximation  
    
    \begin{eqnarray}
        \frac{a_1}{R_\mathrm{in}} = 0.2 g^{1/3} \fracbrac{K}{10}^{1/6} \fracbrac{m_1}{10 m_\oplus}^{1/3} 
        \left( 1 + \frac{m_1}{m_2} \right)^{-1/6} 
        \fracbrac{0.01 M_\odot}{M_\mathrm{disk}}^{1/3}
         \left( \frac{R_\mathrm{in}}{R_\mathrm{out}} (1 + \frac{R_\mathrm{in}}{R_\mathrm{out}} ) \right)^{-1/3}
        (1 - \alpha^{3/2})^{-1/3}
    \end{eqnarray}    

    If the disk is closer than this to the planet our approximation will significantly underestimate the differential precession rate, and equation \ref{eq:precham} must be used to obtain the true differential rate.  
    Finally, we wish to briefly consider sources of precession other than massive protoplanetary disks. Our Hamiltonian model shown in Section~\ref{sec:dynmodel} is agnostic to the source of precession, therefore it is straightforward to consider other sources.
    A number of studies have considered the role of the time-varying quadrupole moment from rapidly spinning young stars on systems' secular dynamics \citep[e.g.,][]{Veras_2007,Spalding_2017,Schultz_2021}.

    Here we consider the influence of a stellar $J_2$ moment on the \emph{resonant} dynamics pair of planets. For low eccentricity planets the precession rate is related to the $J_2$ by         
    \begin{equation}
        \frac{\dot{\varpi}}{n_p} = \frac{3}{2} J_2\fracbrac{R_*}{a_p}^2,
        \label{eq:oblateprec}
    \end{equation}

    where  $R_*$ is the radius of the star , $a_p$ the semi-major axis of the planet, and $n_p$ the mean motion of the planet~\citep[e.g.,][]{Greenberg_1981}.  Studies of young, quickly rotating stars suggest these values of $J_2$ could rise as high as $10^{-2}$ in some systems \citep{Zahn_2010}.  Equation \eqref{eq:oblateprec} implies that short period planets may have large precession rates $\dot{\varpi}/n_p \approx 10^{-3}$    , which are comparable to those generated from a massive protoplanetary disks.  Since the dependence of the precession rate on the distance is steep, the differential precession rate between two planets orbiting around such a star to be large.  This implies that the innermost planets in such a system will have a much higher precession rate than the outer, resulting in a large differential precession rate.  
    
        \begin{eqnarray}
            \frac{a_1}{R_*} = 3.7 f^{-1/2} \fracbrac{J_2}{10^{-3}}^{1/2} \fracbrac{10}{K}^{1/4} \fracbrac{M_*}{M_\odot}^{1/2} 
            \fracbrac{10 M_\oplus}{m_2}^{1/2} \left( 1 + \frac{m_1}{m_2} \right)^{1/4} \alpha^{-5/8} (1 - \alpha^{7/2})^{1/2}
        \end{eqnarray}

        A quadrupole potential felt by a planet around a circumbinary system can also be approximated with equation \ref{eq:oblateprec}, in this case $J_2 = \frac{1}{2(1-e_B)^{3/2}}q(1-q)$ where $q=M_2/(M_1+M_2)$ and $e_B$ is the binary eccentricity, and the stellar radius is set equal to the semi-major axis of the binary $a_B$.In practice, studies of circumbinary planets \citep[e.g,][]{Leung_2013} find precession timescales as low as $50$ years, with corresponding precession rates as high as $\dot{\varpi}/n_p \approx 10^{-2}$.
        Since the precession rate has a steep radial dependence, large differential precession rates will also occur in these systems. The differential precession rates in both these types of systems are comparable to the rates in \ref{fig:diskprecession} and may be large enough to cause changes in equilibrium eccentricity of any orbiting resonant planets. 
        
        Future areas of study might include efforts to generalize our model to planets with higher eccentricities or to include the change in mean-motion induced by a massive disk. Both of these efforts would help to better understand the effect of disk precession on resonant capture.  Additionally, it may be fruitful to attempt detailed, hydrodynamical modeling of planets embedded in their disks.  Such efforts could help to better characterize the precession rates of embedded planets and could shed light on how far from equilibrium such systems will capture.

\section{Summary}
\label{sec:summary}
    In this paper, we show how differential precession between the two planets can cause deviations in the captured equilibrium eccentricities away from their precession-free values. We show that resonance capture can excite extreme eccentricities when differential precession is sufficiently strong and when the eccentricity damping felt by the more slowly precessing body is small. We argued that this situation could arise when a resonant planet pair migrates into the inner cavity of a protoplanetary disk through planet-disk interactions.  We show these bodies exhibit two main types of behaviors depending on whether the differential precession is above or below a critical value. 
    
    More generally, for planets captured in a protoplanetary disk, rapid dispersal of the disk will strand planets away from their precession-free equilibria. This will induce large oscillations in eccentricity if the differential precession rate is smaller than the critical value.  For systems where the differential precession rate is larger than the critical value, the rapid decrease in differential precession can destabilize the planets.    

    This work demonstrates that differential precession has a significant impact on resonance capture. This more detailed understanding of resonance capture including the differential precession induced by the natal disk may prove necessary for interpreting the growing population of well characterized exoplanet systems.

\bibliography{refs}{}
\bibliographystyle{aasjournal}
\end{document}